\pdfoutput=1
\PassOptionsToPackage{sort}{natbib}
\documentclass[11pt]{article}

\usepackage[final]{acl}

\usepackage{times}
\usepackage{latexsym}

\usepackage[T1]{fontenc}

\usepackage[utf8]{inputenc}

\usepackage{microtype}

\usepackage{inconsolata}

\usepackage{graphicx}

\usepackage{import}
\usepackage{graphicx}
\usepackage{amsmath}
\usepackage{amssymb}
\usepackage{mathbbol}
\usepackage{bm}
\usepackage{booktabs}
\usepackage{float}
\usepackage{listings}
\usepackage{multirow}
\usepackage{lipsum}  
\usepackage{subcaption}
\usepackage{comment}
\usepackage[switch]{lineno}
\usepackage{wrapfig}
\usepackage{varwidth}
\usepackage{array}
\usepackage{makecell}
\usepackage[capitalize]{cleveref}
\usepackage{todonotes}
\usepackage[normalem]{ulem}
\usepackage{soul}
\useunder{\uline}{\ul}{}

\usepackage{minted}

\usepackage{pifont}

\newcommand{\rinlinecode}[1]{\lstinline[basicstyle=\ttfamily, breaklines=true, breakatwhitespace=true]|#1|}

\newcommand{\marco}{MS~MARCO}

\definecolor{promptcolor}{HTML}{EEEEEE}

\sethlcolor{promptcolor}
\newcommand{\bc}[1]{\hl{#1}}

\newcommand{\FastResultHeapq}{\bc{FastResultHeapq}}\newcommand{\MaterializedQRel}{\bc{MaterializedQRel}}\newcommand{\MaterializedQRelConfig}{\bc{MaterializedQRelConfig}}\newcommand{\BinaryDataset}{\bc{BinaryDataset}}\newcommand{\DataArguments}{\bc{DataArguments}}\newcommand{\EncodingDataset}{\bc{EncodingDataset}}\newcommand{\MultiLevelDataset}{\bc{MultiLevelDataset}}\newcommand{\RetrievalCollator}{\bc{RetrievalCollator}}\newcommand{\EvaluationArguments}{\bc{EvaluationArguments}}\newcommand{\IRMetrics}{\bc{IRMetrics}}\newcommand{\RetrievalEvaluator}{\bc{RetrievalEvaluator}}\newcommand{\BiEncoderRetriever}{\bc{BiEncoderRetriever}}\newcommand{\ModelArguments}{\bc{ModelArguments}}\newcommand{\PretrainedEncoder}{\bc{PretrainedEncoder}}\newcommand{\PretrainedRetriever}{\bc{PretrainedRetriever}}\newcommand{\RetrievalLoss}{\bc{RetrievalLoss}}\newcommand{\RetrievalTrainer}{\bc{RetrievalTrainer}}\newcommand{\RetrievalTrainingArguments}{\bc{RetrievalTrainingArguments}}

\DeclareRobustCommand{\trovelogo}{\begingroup\normalfont
  \vspace{-0.2em}\raisebox{-0.4em}{\includegraphics[height=1.5em]{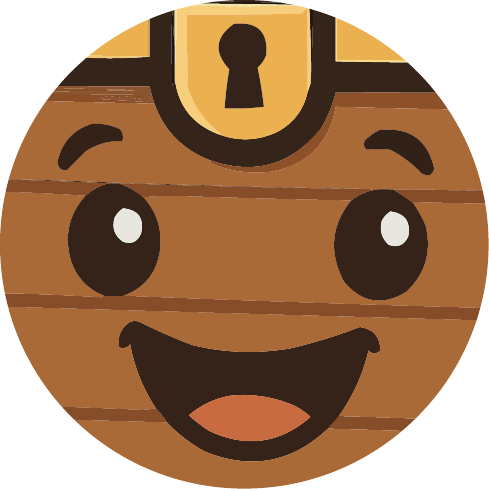}}\kern 0.4em\endgroup
}

\title{Trove: A Flexible Toolkit for Dense Retrieval}

\author{
\textbf{Reza Esfandiarpoor} \quad
\textbf{Max Zuo} \quad
\textbf{Stephen H. Bach}
\\
Department of Computer Science, Brown University \\
\texttt{\{reza\_esfandiarpoor,zuo,stephen\_bach\}@brown.edu} \\
Demo: {\tt \hypersetup{urlcolor=magenta}\href{https://youtu.be/rThGHOw3wS8}{https://youtu.be/rThGHOw3wS8}} \\
{\tt \trovelogo \hypersetup{urlcolor=magenta}\href{https://ir-trove.dev}{ir-trove.dev}}
}

\begin{document}
\maketitle
\begin{abstract}

We introduce Trove, an easy-to-use open-source retrieval toolkit that simplifies research experiments without sacrificing flexibility or speed.
For the first time, we introduce efficient data management features that load and process (filter, select, transform, and combine) retrieval datasets on the fly, with just a few lines of code.
This gives users the flexibility to easily experiment with different dataset configurations without the need to compute and store multiple copies of large datasets.
Trove is highly customizable: in addition to many built-in options, it allows users to freely modify existing components or replace them entirely with user-defined objects.
It also provides a low-code and unified pipeline for evaluation and hard negative mining, which supports multi-node execution without any code changes.
Trove's data management features reduce memory consumption by a factor of 2.6.
Moreover, Trove's easy-to-use inference pipeline incurs no overhead, and inference times decrease linearly with the number of available nodes.
Most importantly, we demonstrate how Trove simplifies retrieval experiments and allows for arbitrary customizations, thus facilitating exploratory research.

\end{abstract}

\section{Introduction}
\label{sec:intro}

Influential toolkits such as those provided by Hugging Face (HF) simplify Machine Learning (ML) pipelines and support extensive customization with minimal effort, thus facilitating exploratory research~\citep{hftransformers,hfdatasets,hfaccelerate}.
Similarly, existing retrieval toolkits have significantly improved Information Retrieval (IR) pipelines~\citep{tevatron,sbert}.
However, IR experiments still require a considerable amount of engineering effort for many tasks like efficient data management or model customization.
Here, we introduce a novel open-source toolkit that simplifies various stages of retrieval pipelines, enabling efficient data management, flexible modeling, and easy distributed evaluation.
Our design prioritizes customization and makes it easy to freely modify or entirely replace each component.

General-purpose toolkits are not directly applicable in retrieval pipelines.
Retrieval is uniquely different from most ML problems in that instances of retrieval tasks are not self-contained.
For example, while solving an image classification task only involves one image, a single retrieval task involves one query and the \emph{entire corpus}.
This makes retrieval experiments more challenging.
Since most data management tools like HF Datasets~\citep{hfdatasets} process each instance isolated from the rest of the dataset, they cannot be directly used in retrieval pipelines~\citep{tevatron}.
Distributed evaluation is also more challenging.
Instances of retrieval tasks share a lot of the computation (i.e., encoding the corpus), and we cannot simply evaluate each instance on a separate device and aggregate the results.
Finally, HF transformers models only provide the encoder, and we cannot directly use them for retrieval without additional modeling.

\begin{figure}[t!]
  \centering
  \includegraphics[width=.9\columnwidth]{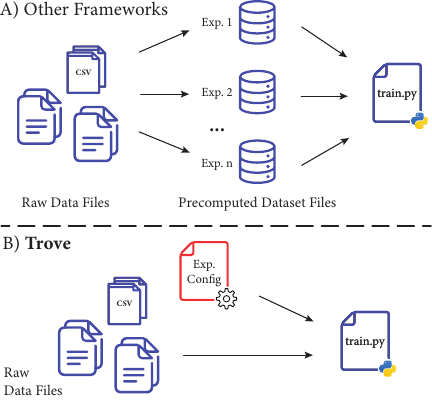}
  \caption{A) Existing toolkits require manually creating and maintaining large pre-processed data files for each experiment. B) Trove processes datasets on the fly based on the given configuration options.}
  \label{fig:data_proc}
  \vspace{-4mm}
\end{figure}

Existing toolkits have recognized these issues and offer initial solutions.
However, these solutions are often not as flexible or easy to use.
Since naive on-the-fly data preparation for retrieval is memory-intensive, current toolkits rely on large pre-processed dataset files~\citep{tevatron}, often duplicating a lot of data for variations of a single dataset (\cref{fig:data_proc} top).
Although evaluating retrieval tasks is computationally more demanding, currently distributed evaluation is either limited to a single node~\citep{mteb,sbert} or involves several steps and more engineering effort~\citep{tevatron}.
For modeling, current frameworks wrap transformers models in fixed classes, and customizations are limited to a set of pre-defined options.
As a result, exploratory experiments require significant engineering effort, which slows down novel research.

In this work, we introduce Trove, an open-source library that simplifies dense retrieval experiments without sacrificing flexibility.
Trove is the first toolkit to provide features for efficiently managing and pre-processing retrieval data on the fly (\cref{fig:data_proc} bottom).
Trove provides a simple interface for multi-\textbf{node}/GPU inference and is fully compatible with HF transformers ecosystem.
Our modeling approach provides direct access to model components and allows arbitrary customizations.
In general, our design increases flexibility at three levels.
1) Trove provides various built-in options to customize experiments.
2) Our transparent design allows users to override many methods with custom logic.
3) Trove's modular structure allows users to entirely replace many components with arbitrary objects.
In summary:

\begin{itemize}
    \item Trove is built around the unique characteristics of the retrieval task and, for the first time, offers fast and memory-efficient operations for loading and pre-processing (filter, transform, combine, etc.) retrieval data \emph{on the fly}.
    \item Trove provides a simple and unified interface for evaluation and hard negative mining, which supports both multi-node and multi-GPU inference without additional code.
    
    \item Trove allows for direct customization of all modeling components or even replacing them with arbitrary modules, while maintaining compatibility with HF transformers ecosystem.
    
    \item Trove is designed with customization in mind.
    The codebase is heavily documented and easy to understand.
    We provide ample guides and examples to facilitate customization.
\end{itemize}

\section{Background and Challenges}
\label{sec:motivation}

There is a growing body of work on transformer-based dense retrievers. 
Many works have focused on improving the training data through techniques like using mined hard negatives or a large number of random in-batch negatives~\citep{zerveas_coder_2022,zerveas-etal-2023-enhancing,karpukhin-etal-2020-dense,qu-2021-rocketqa,moreira2024nv,xiong2020approximate,zhan2021optimizing,rekabsaz_tripclick_2021}.
Several works have also used synthetic data for training~\citep{li2024syneg,dai2022promptagator, bonifacio2022inpars, jeronymo2023inpars, alaofi2023can,lee2024geckoversatiletextembeddings}.
With the introduction of RepLLaMA~\citep{ma2024fine}, large decoder-only LLMs and PEFT techniques~\citep{hu2022lora} have become popular for retrieval models~\citep{wang-etal-2024-improving-text}.
There are also new variants of the retrieval task itself, such as retrieval instructions~\citep{weller2024promptriever,asai2022task}.

To illustrate the challenges of developing IR pipelines, we compare the development experience for IR to other ML tasks for three common operations.

\noindent\textbf{Data Management}\hphantom{A}
For most ML problems, tools like HF Datasets load, pre-process, and combine multiple datasets on the fly with just a few lines of code and very little memory overhead.
By contrast, for common IR datasets like \marco{}~\citep{bajaj_msmarco_2018}, just creating training samples from raw dataset files requires a lot of memory and extra code.
Existing tools do not offer any data management functionality and instead rely on large pre-processed data files, which require maintaining many large files with duplicate data.
In addition to the cumbersome process, with this approach, data changes are not trackable by VCS, which hurts reproducibility.

\noindent\textbf{Modeling}\hphantom{A}
Usually, users have full control over the model and can apply \emph{arbitrary} customizations (e.g., add LoRA adapters or change the loss).
However, current retrieval toolkits wrap the encoder in custom classes, without providing direct access to the transformers backbone.
As a result, any customization requires explicit support from the library\footnote{Example: \href{https://github.com/UKPLab/sentence-transformers/issues/2575}{sentence-transformers/issues/2575}}.
The interactions between different components also limit flexibility.
For example, it is not possible to train with graduated relevance labels using Tevatron~\citep{tevatron} even if we overwrite the loss function.

\noindent\textbf{Inference}\hphantom{A}
The evaluation pipeline often involves a simple script that executes the forward pass and calculates the metrics, all in one job step.
For distributed evaluation, users just need to execute the same script with a distributed launcher like Accelerate~\citep{hfaccelerate}, with minimal changes.
Although IR evaluation is computationally more demanding, multi-node execution is not straightforward.
The evaluation process with SentenceTransformers and MTEB\footnote{\href{https://github.com/embeddings-benchmark/mteb}{gh/embeddings-benchmark/mteb}} packages is easy but limited to only a single node.
Tevatron supports multi-node evaluation, but it needs to manually launch multiple jobs to encode each dataset shard separately, and then launch another job to retrieve related documents and another job to calculate the metrics.

\section{System Design}
\label{sec:design}

\begin{figure}[t!]
  \centering
  \includegraphics[width=\columnwidth]{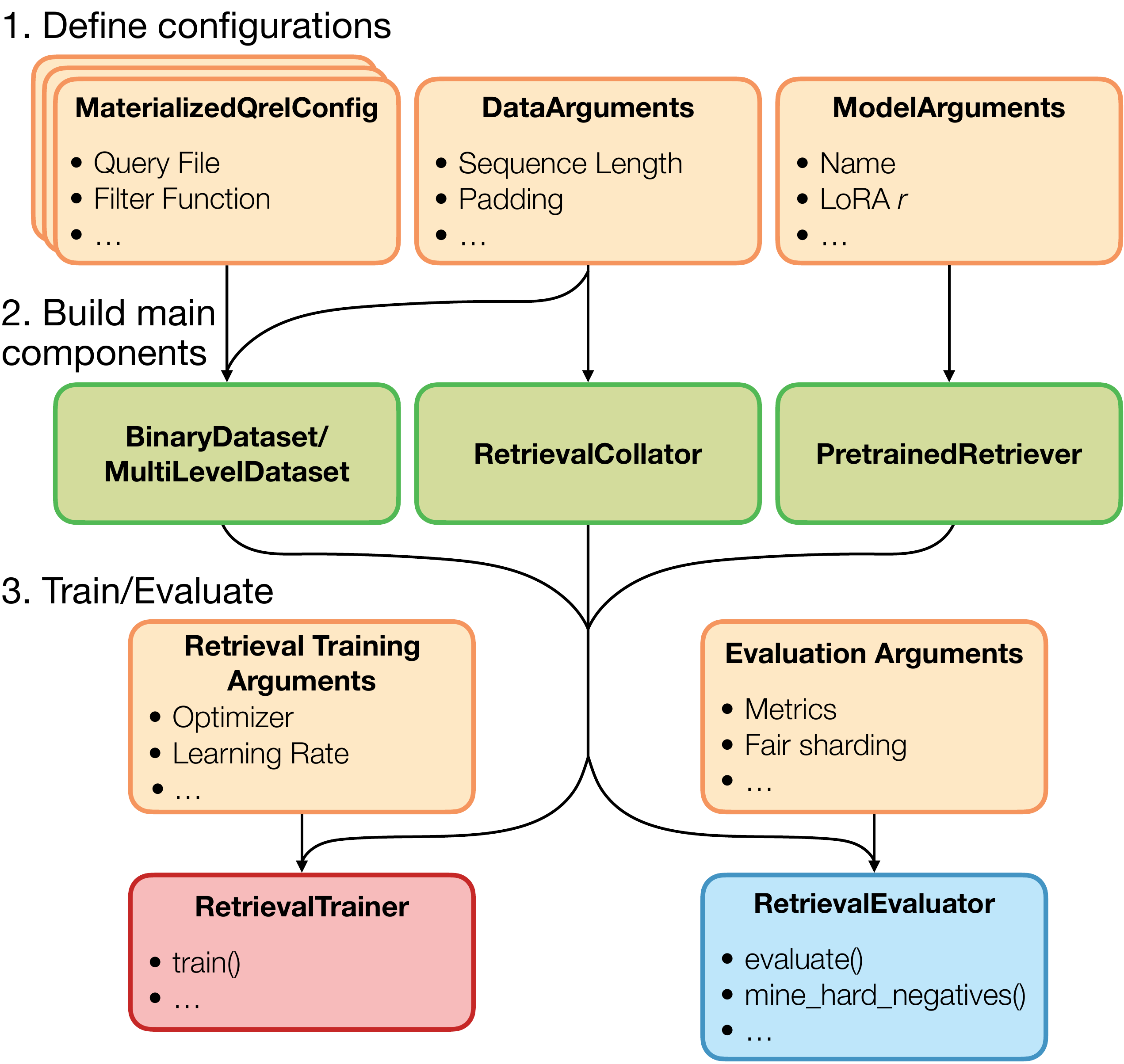}
  \caption{Training and evaluation workflow with Trove.}
  \label{fig:workflow}
\vspace{-4mm}
\end{figure}

Here, we first explain Trove experiment workflows (\cref{fig:workflow}) and then describe its major components.

\subsection{Workflow}

Trove experiment workflows are simple and based on configuration objects.
Users create one or more instances of \MaterializedQRelConfig{} to specify how raw input files (query, corpus, qrels) should be loaded and processed (e.g., filtered).
We also create a \DataArguments{} object with dataset-level details (e.g., sequence length).
Users then use these objects to instantiate one of the main dataset classes (\BinaryDataset{} or \MultiLevelDataset{}).
Next, we create a \ModelArguments{} instance with model details like name, pooling type, and LoRA configuration.
We instantiate the main retriever (e.g., \BiEncoderRetriever{}) from the argument object.
Finally, we use these components in addition to an instance of \RetrievalTrainingArguments{} (\EvaluationArguments{}) to instantiate \RetrievalTrainer{} (\RetrievalEvaluator{}), which is responsible for the main training (evaluation) loop.
Our design allows instantiating configuration objects from command-line arguments, which makes it easier to run diverse experiments.
Moreover, our design increases flexibility by exposing the main components of the pipeline: users can easily customize or entirely replace each module without changing Trove's source code or even major changes to their experiment workflow.

\subsection{Data Management}

Large IR datasets are often made of three sets of files: query, corpus, and qrels (i.e., annotations).
Creating training instances is simple: for each query, find the related document IDs from the qrel files and then replace these IDs with the content from the query and corpus files.
Efficiently implementing this logic is challenging for large datasets like \marco{} with 500K queries and 8M documents.
Just loading all the query and corpus records consumes a lot of memory.
Moreover, with multiple datasets or more complex pipelines, pre-processing or merging millions of qrel records slows down the program.

\subsubsection{Internal Representation}
We implement \MaterializedQRel{}, an efficient container for IR data that holds query, corpus, and qrel records.
We use the Polars library\footnote{\url{https://pola.rs}} to efficiently group qrel triplets by query ID, which significantly speeds up pre-processing and lookup operations for related documents.
We convert query, corpus, and grouped qrel records to memory-mapped Apache Arrow tables that are indexable by ID.
\MaterializedQRel{} only works with IDs, \emph{without loading the actual data}.
For each training instance, we load the data at the very last step and even then only load the necessary records for the current instance.
As a result, \MaterializedQRel{} significantly reduces memory consumption.

\MaterializedQRel{} enables users to process the data just by setting a few config values.
For instance, users can easily filter qrel triplets, select a subset of queries, or change the labels.
See~\cref{sec:demo} for examples.

\subsubsection{User-facing Classes}

These benefits are available to users through the \MultiLevelDataset{} and \BinaryDataset{} classes.
Trove datasets are made of one or more \MaterializedQRel{} instances, which allows defining complex data pipelines.
For instance, users can apply different pre-processing steps to each data source (e.g., real and synthetic) before combining them.
To create a dataset, users just need to initialize the dataset class with config objects that specify data loading and processing details.
As a result, Trove's data pipelines are trackable with VCS.

We also implement \EncodingDataset{}, which prepares the data for encoding during inference and simplifies embedding cache management.
We implement the embedding cache as memory-mapped Apache Arrow tables that are indexable by ID.
The embedding cache supports \emph{lazy loading} and only loads each cached vector when it is needed.
The interface is simple: users call \rinlinecode{cache\_records(ids, vectors)} to write the vectors and their IDs to cache.
Later, when accessing each item (i.e., \rinlinecode{dataset[i]}), the dataset returns the cached embedding instead of raw text if available.
Additionally, we implement the \RetrievalCollator{}, which is responsible for tokenizing and batching examples for retrieval.

\subsubsection{Additional Optimizations}

\noindent\textbf{Callbacks for Flexibility}\hphantom{A}
We implement frequently changing operations as callback functions for easier customization.
For example, users can load qrel triplets from custom file formats by registering a loader with \texttt{@register\_loader}.
Users can customize input formatting (e.g., add instructions) by passing \texttt{format\_query} and \texttt{format\_passage} callbacks to the dataset.
Similarly, the \texttt{filter\_fn} option in \MaterializedQRelConfig{} allows users to filter the qrel triplets with custom functions.

\noindent\textbf{Reliability}\hphantom{A}
We cache intermediate artifacts in the first run and track changes using a fast fingerprinting method.
We also use atomic write operations to guard against corrupted files.
As a result, datasets are very fast after the first run and reliably generate the same data in all runs.

\subsection{Modeling}
\label{sec:design_modeling}
Trove's modeling is divided into three main components (retriever, encoder, and loss) and allows users to customize each component independently.

\noindent\textbf{Retriever}\hphantom{A}
Subclasses of \PretrainedRetriever{} are the main model class in Trove and consist of an encoder, loss function, and the retrieval logic.
\PretrainedRetriever{} can use all HF transformers models as encoder and supports common pooling and normalization operations, as well as LoRA adapters and quantization.
It provides the \rinlinecode{from\_model\_args()} method that creates the correct encoder and loss function based on the given options.
To allow arbitrary customizations, we encapsulate all details related to transformers models (e.g., quantization) in the encoder and allow users to use arbitrary \rinlinecode{nn.Module} objects as the encoder.

Users can subclass \PretrainedRetriever{} and overwrite the \rinlinecode{forward()} method to implement custom retrieval logics.
Trove already comes with \BiEncoderRetriever{}, which implements the dual-encoder retrieval logic with support for cross-device in-batch negatives.

\noindent\textbf{Encoder}\hphantom{A}
To experiment with new encoding methods (e.g., different pooling or PEFT techniques), users can implement custom encoder wrappers as \PretrainedEncoder{} subclasses.
Compared to using arbitrary \rinlinecode{nn.Module} objects as encoder, this allows us to swap encoder wrappers without changing the code, which simplifies user scripts.
Users just need to instantiate the retriever with different options (e.g., \rinlinecode{encoder\_class="MyEncoderClass"}).

\noindent\textbf{Loss Function}\hphantom{A}
Trove implements the InfoNCE and KL Divergence losses.
Users can implement new loss functions as \RetrievalLoss{} subclasses and choose the correct loss through retriever options (e.g., \rinlinecode{loss="MyLossClass"} or \rinlinecode{"kl"}).

\begin{figure*}[t]
\centering
\begin{minted}[mathescape,
           numbersep=5pt,
           autogobble,
           frame=lines,
           fontsize=\small,
           breaklines,
           tabsize=2,
           linenos,
           framesep=2mm]{python} 
from transformers import AutoTokenizer, HfArgumentParser
from trove import *

parser = HfArgumentParser((RetrievalTrainingArguments, ModelArguments, DataArguments))
train_args, model_args, data_args = parser.parse_args_into_dataclasses()

tokenizer = AutoTokenizer.from_pretrained(...)
model = BiEncoderRetriever.from_model_args(...)
collator = RetrievalCollator(data_args, tokenizer, append_eos=False)

pos = MaterializedQRelConfig(min_score=1, qrel_path="qrels/train.tsv", ...)
neg = MaterializedQRelConfig(group_random_k=2, qrel_path="mined_neg.tsv", ...)
dataset = BinaryDataset(data_args, model.format_query, model.format_passage, pos, neg)

trainer = RetrievalTrainer(model, train_args, collator, dataset)
trainer.train()
\end{minted}
\caption{Training with Mined Hard Negatives}
\label{code:train_bin}
\vspace{-4mm}
\end{figure*}

\subsection{Training}

Inspired by Tevatron~\citep{tevatron}, we ensure all Trove components are compatible with HF transformers and directly use its Trainer module for training, with minimal changes.

Trove makes it possible to approximate IR metrics like nDCG during training by ranking a small number of annotated documents for each development query, similar to a reranking task.
We introduce \IRMetrics{}, which can be used as the \rinlinecode{compute\_metric} callback to efficiently calculate approximate IR metrics during training for small instances of \MultiLevelDataset{}.

\subsection{Inference}

\RetrievalEvaluator{} class implements a simple and unified interface for evaluation and hard negative mining.
Inference is as easy as creating an instance of \RetrievalEvaluator{} and calling the \rinlinecode{evaluate()} or \rinlinecode{mine\_hard\_negatives()} method.
Trove supports logging to Wandb and can integrate other experiment trackers using callback functions.

For distributed inference, we just need to launch the same script, without any changes, using a distributed launcher.
\RetrievalEvaluator{} automatically distributes the computation across available \emph{nodes} and GPUs.
We also introduce a \emph{fair sharding} feature that allows mixing GPUs with different capabilities without stalling the faster devices.
Trove adjusts the shard sizes based on GPU throughput, assigning more samples to faster devices.

Trove introduces \FastResultHeapq{}, a Pytorch alternative to naive Python heapq that uses fast matrix operations and GPU acceleration for tracking the top-k documents for each query\footnote{This is not a full heapq. It just mimics some functionalities to keep track of topk documents for each query.}.
Existing frameworks commonly use Python's heapq, which is a major bottleneck and stalls GPU cycles during evaluation~\citep{mteb}.
\FastResultHeapq{} is 16x and 600x faster than Python heapq for online and cached embeddings, respectively.

\section{Demonstration}
\label{sec:demo}

Here, we demonstrate Trove's flexibility and ease of use and benchmark its efficiency.

\subsection{Flexibility and Ease of Use}

We have already used Trove for large-scale research experiments in our earlier work, SyCL~\citep{sycl}.
Below, we outline the pipeline for two key SyCL experiments. Trove can be easily installed from PyPI:

\begin{minted}[mathescape,
           frame=none,
           fontsize=\small,
           breaklines,
           tabsize=2,
           framesep=0mm]{bash} 
        $ pip install ir-trove
\end{minted}

Trove greatly reduces the engineering effort required for common training setups.
\Cref{code:train_bin} shows the complete code needed to train dense retrievers with mined hard negatives using InfoNCE loss.
This simple code already supports multi-node/GPU training, standard pooling and normalization operations, LoRA adapters, and quantization.

Now, we modify the code to train on a mix of multi-level synthetic data (labels in \{0, 1, 2, 3\}), annotated positives, and mined hard negatives. To do this, we simply replace lines 11–13 in~\cref{code:train_bin} with the following:

\begin{minted}[mathescape,
           autogobble,
           frame=none,
           fontsize=\small,
           breaklines,
           tabsize=2,
           framesep=0mm]{python} 
syn = MaterializedQRelConfig(...,
    qrel_path="synth_qrel.tsv",
    corpus_path="synth_corpus.jsonl",
    query_subset_path="qrels/orig_train.tsv")
pos = MaterializedQRelConfig(...,
    qrel_path="qrels/train.tsv",
    score_transform=3,
    min_score=1)
neg = MaterializedQRelConfig(...,
    qrel_path="mined_neg.tsv",
    score_transform=1,
    group_random_k=2)
dataset = MultiLevelDataset([syn, pos, neg], ...)
\end{minted}

\noindent This snippet processes each data source differently and combines the results.
\texttt{syn} collection selects only synthetic data for training queries, using query IDs from \rinlinecode{qrels/train.tsv} file.
\texttt{pos} collection filters for documents with relevance labels $\geq 1$ (i.e., positives), then assigns them a new label of 3.
And, \texttt{neg} randomly selects two of the hard negatives per query and assigns them a new label of 1.

In SyCL, we also explore the Wasserstein distance as loss function.
For this, we just implement the loss function as the following and use \rinlinecode{--loss=ws} to run the training script.

\begin{minted}[mathescape,
           autogobble,
           frame=none,
           fontsize=\small,
           breaklines,
           tabsize=2,
           framesep=0mm]{python} 
class WSLoss(RetrievalLoss):
    _alias = "ws"
    
    def forward(self, logits, label):
        loss = ... # calculate the loss value
        return loss
\end{minted}

For training results and additional experiments using Trove, we refer readers to the SyCL paper and codebase\footnote{\href{https://github.com/BatsResearch/sycl}{gh/BatsResearch/sycl}}.
Also, see~\cref{sec:model_customization} for examples that customize Trove's built-in models.

\subsection{Efficiency}

Here, we benchmark the impact of Trove’s optimizations for data management and inference.

\begin{table}[t]
\centering
\centering
\resizebox{0.9\columnwidth}{!}{
\begin{tabular}{@{}lcc|cc@{}}
\toprule
& \multicolumn{2}{c}{Real}
& \multicolumn{2}{c}{Real w/ Synth.} \\

\cmidrule(lr){2-3}
\cmidrule(lr){4-5}

& 1x GPU & 8x GPU & 1x GPU & 8x GPU \\

\cmidrule(lr){2-5}

Naive & 8.85 & 70.80 & 11.30 & 90.40 \\
Trove & \textbf{3.34} & \textbf{26.72} & \textbf{4.07} & \textbf{32.56} \\ \bottomrule

\end{tabular}
}
\caption{Memory consumption in GB}
\label{tab:mem}
\vspace{-4mm}
\end{table}

\noindent\textbf{Data Management}\hphantom{A}
\Cref{tab:mem} shows the memory required to prepare \marco{} data for training.
The naive baseline loads the entire data in memory.
Trove cuts memory usage by 2.6x.
When combining synthetic and real data~\citep{sycl}, it uses only 0.73\,GB of extra memory for loading the additional 2M synthetic passages, far less than the 2.45\,GB required by the naive approach.
This efficiency is critical for distributed training, where each process loads its own data.
On a machine with 8 GPUs, the naive method consumes 90\,GB of RAM just for data loading.
Note that, unlike other frameworks, Trove processes the data on the fly and does not rely on large pre-processed files for each experiment.

\Cref{tab:ttfs} in the Appendix reports the time to first sample (TTFS), which measures the time required to load and process the data.
Thanks to Trove's internal caching, after the first run, the data is available almost instantaneously.
While TTFS has minimal impact on long-running experiments, a short TTFS is critical for efficient debugging and interactive development.

\noindent\textbf{Inference}\hphantom{A}
\Cref{tab:inf_time} shows retrieval times for all queries in \marco{} using E5-Mistral-Instruct~\citep{wang-etal-2024-improving-text} in a distributed environment.
Inference time decreases linearly with the number of nodes, showing that Trove uses additional nodes with no overhead.
Crucially, we just need to run the same script with a distributed launcher, without changing the code.

\begin{table}[t]
\centering
\begin{tabular}{@{}lccc@{}}
\toprule
                 & 1x Node & 2x Node & 3x Node \\ \cmidrule{2-4}
Inference Time    & 14:20  & 7:12 & 4:48 \\
\bottomrule
\end{tabular}
\caption{Inference time in HH:MM format for different number of nodes}
\label{tab:inf_time}
\end{table}

\begin{table}[t]
\centering
\centering
\resizebox{\columnwidth}{!}{
\begin{tabular}{@{}lcc|cc@{}}
\toprule
\multicolumn{3}{l}{Queries \hphantom{AA} On The Fly}
& \multicolumn{2}{c}{w/ Cached Embs} \\

\cmidrule(l){1-5}

& Naive & Trove & Naive & Trove \\

\cmidrule(lr){2-3}
\cmidrule(lr){4-5}

6K & 1h:9m & \textbf{7s} & 21s & \textbf{1s} \\
500K & 130h:40m & \textbf{11m:45s} & 30m:17s & \textbf{1m:52s} \\
\bottomrule

\end{tabular}
}
\caption{Performance of Python's heapq vs Trove's \FastResultHeapq{} during inference}
\label{tab:hq}
\vspace{-4mm}
\end{table}

\Cref{tab:hq} compares the performance of Python's heapq with \FastResultHeapq{} for keeping track of top-k documents at \marco{} scale.
In an online setup where we embed a small batch of 256 documents and compare it with queries on the fly, Trove is more than 600x faster than Python's heapq.
When the number of queries grows (e.g., for hard negative mining), Python's heapq becomes unusable, taking up to 130 hours.

Even when embeddings are cached and comparisons are made in large batches (e.g., 40,960 documents) on GPU, Trove remains 16x to 21x faster.
However, in practice, this speedup is not realized for Python's heapq.
It is often bottlenecked by disk I/O, particularly with the simple caching mechanisms used by existing frameworks.
\section{Conclusion}

In this work, we introduce Trove, an open-source toolkit for dense retrieval that reduces the engineering effort in research experiments.
Trove eliminates the need for large pre-processed data files and, for the first time, provides data management features that load and process retrieval data on the fly, with a small memory footprint.
Trove provides full control over modeling and allows users to freely customize different modeling components.
Trove provides a simple and unified interface for evaluation and hard negative mining, which supports multi-node inference without any extra code.
While Trove provides a simple high-level interface, every component is designed to be configured, modified, or replaced entirely.
As a result, Trove provides researchers with a tool to quickly and freely experiment with new ideas.

\section*{Acknowledgements}

This material is based upon work supported by the National Science Foundation under Grant No. RISE-2425380. Any opinions, findings, and conclusions or recommendations expressed in this material are those of the author(s) and do not necessarily reflect the views of the National Science Foundation. This research is supported in part by the Office of Naval Research (ONR) award N00014-20-1-2115. Disclosure: Stephen Bach is an advisor to Snorkel AI, a company that provides software and services for data-centric artificial intelligence.

\bibliography{misc/refs}

\appendix

\section{Limitations}

Since Trove is mainly aimed at researchers, our design emphasizes a simple codebase that is freely customizable.
As a result, Trove does not support all retrieval-related methods out of the box, but makes it easy to implement arbitrary methods in user scripts.
On the other end of the spectrum, there is SentenceTransformers, a great library that makes it easy to get started with IR experiments, with built-in support for many retrieval methods.
Inevitably, such a codebase is complex and hard to modify by users, which is what we are trying to avoid with Trove.

Moreover, to facilitate exploration and rapid experiments, we sometimes avoid industry standards and implement our own solutions.
For example, we implement a fast embedding cache and our own Pytorch container (i.e., \FastResultHeapq{}) to speed up nearest neighbor search instead of using tools such as FAISS~\citep{faiss}, which are also used by other libraries like Tevatron.
While FAISS has many great features, in our case, creating a large search index that is only used once is not as efficient.
Moreover, such dependencies limit flexibility.
For example, FAISS only reports the similarity scores of the most similar documents for each query.
On the other hand, our method can also track similarity scores for arbitrary documents even if they are not ranked among top-k results but, for example, are useful for answering a specific research question.

\begin{table}[h]
\centering
\centering
\resizebox{0.9\columnwidth}{!}{
\begin{tabular}{@{}lcc|cc@{}}
\toprule
& \multicolumn{2}{c}{Real}
& \multicolumn{2}{c}{Real w/ Synth.} \\

\cmidrule(lr){2-3}
\cmidrule(lr){4-5}

& TTFS & TTFS\textsubscript{ 1st} & TTFS & TTFS\textsubscript{ 1st} \\

\cmidrule{2-5}

Naive & 31 & 31 & 40 & 40 \\
Trove & 5 & 39 & 7 & 55 \\
\bottomrule
\end{tabular}
}
\caption{Time to first sample (TTFS) in seconds for the first and subsequent runs}
\label{tab:ttfs}
\end{table}

\section{Model Customization}
\label{sec:model_customization}

Trove provides different methods for customizing the model.
As described in~\cref{sec:design_modeling}, Trove comes with many built-in options for choosing different pooling operations, applying embedding normalization, adding LoRA adapters, or quantization.
Here, we provide several examples of how users can further customize models beyond these options.

\paragraph{Input Formatting}
For convenience, Trove encoders include two methods for proper input formatting for a given model.
The code below, modifies these methods to add instructions to input queries and passages, similar to~\citet{wang2023improving}.

\begin{minted}[mathescape,
           autogobble,
           frame=none,
           fontsize=\small,
           breaklines,
           tabsize=2,
           framesep=0mm]{python} 
class EncoderWithInstructions(DefaultEncoder):
    _alias = "encoder_with_inst"

    def format_query(self, text, dataset, **kwargs):
        if dataset == "msmarco":
            inst = "Instruct: Given a web search query, retrieve relevant passages that answer the query\nQuery: "
        else:
            inst = "Query: "
        return inst + text

    def format_passage(self, text, title=None, **kwargs):
        return f"Passage: {title} {text}"
\end{minted}
Then, in the user scripts (e.g.,~\cref{code:train_bin}), they just need to pass \rinlinecode{--encoder_class="encoder_with_inst"} to use this modified encoder wrapper.

\paragraph{Pooling Method}
Here is an example of how users can modify the default encoder wrapper to implement different pooling operations.

\begin{minted}[mathescape,
           autogobble,
           frame=none,
           fontsize=\small,
           breaklines,
           tabsize=2,
           framesep=0mm]{python} 
class EncoderWithNewPooling(DefaultEncoder):
    _alias = "encoder_new_pooling"
    
    def encode(self, inputs) -> torch.Tensor:
        output = self.model(**inputs, return_dict=True)
        embs = ... # custom pooling and normalization operations
        return embs
\end{minted}
Similar to above, users can use \rinlinecode{--encoder_class="encoder_new_pooling"} to use the above encoder wrapper.

\paragraph{New Encoder Wrappers}
Users can also directly subclass \rinlinecode{PretrainedEncoder}, which gives them control over how the model is loaded, saved, and used for calculating the embeddings.
\begin{minted}[mathescape,
           autogobble,
           frame=none,
           fontsize=\small,
           breaklines,
           tabsize=2,
           framesep=0mm]{python} 
class CustomEncoder(PretrainedEncoder):
     _alias = 'custom_encoder'
    
    def __init__(self, args: trove.ModelArguments, **kwargs):
        self.model = AutoModel.from_pretrained(args.model_name_or_path)
        # Arbitrary Model Customizations (LoRA, Quantization, etc.):
        # ...
        
    def save_pretrained(self, *args, **kwargs):
        ...
        
    def encode_query(self, inputs):
        ...
    
    def encode_passage(self, inputs):
        ...
\end{minted}
These new encoder wrappers are also automatically registered and available through configuration options (e.g., \rinlinecode{--encoder_class="custom_encoder"}).

\paragraph{User-provided Encoder Objects}
Users can also directly instantiate \BiEncoderRetriever{} with any \rinlinecode{nn.Module} object as the encoder.
\begin{minted}[mathescape,
           autogobble,
           frame=none,
           fontsize=\small,
           breaklines,
           tabsize=2,
           framesep=0mm]{python} 
custom_encoder: torch.nn.Module = ... # any encoder module
model = BiEncoderRetriever(encoder=custom_encoder, model_args)
\end{minted}

\end{document}